\pdfoutput=1
\documentclass[10pt]{article}

\usepackage{color}
\usepackage{bbm}
\usepackage{epsfig}
\usepackage{amssymb,amsmath,amscd, mdwmath}
\usepackage{graphicx,cite,url}
\usepackage{algorithmic}
\usepackage{float}
\usepackage{amsfonts}
\usepackage{syntonly}
\usepackage{multirow}
\usepackage{graphicx}
\usepackage{subfig}
\usepackage{tabularx}
\usepackage{setspace}
\usepackage{stfloats}
\usepackage{amsthm}
\usepackage[labelfont=bf,labelsep=period,justification=raggedright]{caption}

\makeatletter
\renewcommand{\@biblabel}[1]{\quad#1.}
\makeatother

\date{}

\begin{document}

% Title must be 150 characters or less
\begin{flushleft}
{\Large
\textbf{A game model for the multimodality phenomena of coauthorship networks}
%\textbf{Modelling  tipping-point phenomena of scientific coauthorship networks     and citation networks}
}
% Insert Author names, affiliations and corresponding author email.
\\
Zheng Xie$^{1, \sharp }$
\\
\bf{1}   College of Science,   National University of Defense Technology, Changsha,   China
\\  $^\sharp$ xiezheng81@nudt.edu.cn
 \end{flushleft}
%Sint-Andriesstraat 2, 2000
% Please keep the abstract between 250 and 300 words
\section*{Abstract}
%复杂网络中的博弈行为与演化机理分析
%Compared with atoms or stars in physical world,  human society is complex, which has a variety of complex structures. Therefore, though scientists have discovered many inherent laws underlying a series of physical phenomena from atomic fusion to star evolution, they know little about the behaviors of human society.
%Do there also exist inherent rules behind the social complexity? We try to answer this question through  scientific collaboration  behaviors. A geometric graph for the evolution is proposed, and is validated against a data set of papers published in PNAS during 2000-2015. The validation shows the ability to reproduce a range of features observed with citation and coauthorship data combined and separately. The model reveals how the decisions of individuals in networks generate a range of complex behaviors, such as scale-free and small-world, and how pass through  the divide between ``$1+1=2$" and ``$1+1>2$".

%Scientific cooperations,   featured  by strong pragmatism,
%are generated by the strategies of
% researchers,   essentially are
%   optimized combinations of human  resources.

We provided a game model to simulate the evolution of coauthorship networks,  a   geometric hypergraph built on a circle.
  The model
  expresses kin selection  and network reciprocity, two typically  cooperative  mechanisms,   through
 a  cooperation  condition  called  positive  benefit-minus-cost.
 The costs are modelled through space distances. The benefits are modelled through geometric zones that depend on node hyperdegree,
 which gives an expression of the cumulative advantage on
   the  reputations of authors.
   Our findings indicate that
the model gives a reasonable fitting to empirical coauthorship  networks on their
degree  distribution,     node clustering, and so on. It reveals two properties of node attractions,
namely node heterogeneity
  and   fading  with the growth of hyperdegrees, can deduce
 the  dichotomy of nodes'  clustering behavior and    assortativity, as well as
  the trichotomy of degree and hyperdegree distributions: generalized Poisson,
power-law and exponential cutoff.

% The good model-data  fitting shows the reasonability
%of the designed   game mechanisms.

%The model  provides an example of   adopt  cooperative game theory to generate network complexity.

%The cost and benefit of cooperations make
% combinatorial, or called cooperative, game theory
%be a possible approach   to
%reveal  the   complexity of   coauthorship  networks.

\section*{Introduction}

%什么是科研合作网络
%科研合作的主体人合作的因素是什么

%, and geographic space now evolve into interest space
%Collaborations  between researchers  contribute not only to the breakthrough achievement  unattainable by individuals\cite{Borner3,Jones},
% but also to the transmission and fusion of knowledge, and hence  they incubate several interdisciplines\cite{Adams,Shrum,Uzzi, Wuchty}.
Analyzing large-scale coauthorship data
  provides a macro-level view of collaboration patterns in
scientific  research, and has become an important topic of social sciences\cite{Glanzel1,Glanzel2,Sarigol,Mali}.
 Coauthorship relationship can be expressed through  a  hypergraph, where nodes represent authors, and hyperedges express  the author groups  per paper.
The simple graph   extracted from the hypergraph  is termed    coauthorship network, where edges are generated  by connecting each pair of the  nodes
belonging to the same  hyperedge.
Empirical  coauthorship networks are featured by  specific    local   (degree assortativity, high clustering) and global features (fat-tail, small-world)\cite{Newman1,Newman2,Newman3,Newman0,Newman4,XieLL}.

For general networks, these features  have been reproduced by a range of important models, such as modeling   fat-tail   through   preferential attachment or cumulative advantage\cite{Barab,Moody,Perc,Tomassini,Wagner,Santos}, modeling   degree assortativity   by connecting two    non-connected nodes that have  similar   degrees\cite{Catanzaro}.
 For coauthorship networks,  can these emerged features be deduced through
cooperative
mechanisms?
The  five typical  mechanisms of cooperations exist in the evolution of  coauthorship networks\cite{Nowak}.
 Coauthoring behavior  often occurs in a research group (kin   selection).
 Cooperation  contributes to achieve breakthroughs and reputations (direct and indirect reciprocity).
 Sociable researchers and famous  research teams are easy to attract collaborators (network reciprocity, group selection).

% ``How did
%cooperative
%behavior evolve?"\cite{sci125}.

   The inhomogeneity of node influences is an alternative explanation for fat-tail\cite{Krioukov1}. Nodes with wider influences are likely to gain more connections.
   The idea    has been adopted   to simulate the evolution of  coauthorship
  networks,
   in which node influences are modelled by attaching   specific   geometric zones to nodes\cite{Xie3,Xie6}.
  The corresponding models,   geometric hypergraphs,   provide   reasonable  model-data fittings.
             The work in Ref.\cite{XieLL}  gives an explanation for   the models' mechanism
    from the perspective  of   cooperative game.
The costs are modelled through  space distances, and the benefits are expressed through node
reputations.   Node
reputations cumulate over time.
The cooperation condition   is positive  benefit-minus-cost.

  We  provided  a game model for coauthorship networks based on   the  cumulative advantage on reputations: supposing reputation to be a function of node hyperdegree (a node's hyperdegree
   is the number of the hyperedges  containing the node).
 The function  has a flexibility to    reproduce  the  trichotomy of      degree and hyperdegree distributions, namely a generalized Poisson
head, a power-law in the middle part, and an exponential tail. In addition, nodes'  clustering behavior and   assortative  behavior are different form  small degree   nodes to large degree nodes.
These dichotomous phenomena are also predicted by the model.
  %The good model-data fitting shows the reasonability of the model.

This paper is organized as follows: The model and   data are described in Section 2 and 3 respectively;  the multimodality phenomena are discussed
   in Section 4 and 5; Conclusions are drawn in Section 6.

\section*{The model}

%科研合作的主体人合作的因素是什么

A  cooperative game  consists  a set of players  and  a
characteristic function specifying the value (benefit-minus-cost) created by  subsets of
players in the game.  Coauthorship networks can be regarded as  the results of such a game.
Researchers    can be viewed as players.    Coauthoring a researcher with high    reputation is a kind of   benefits, contributing to  academic success.
The  investments of  manpower and material resources
 on   research  are the  costs of cooperations.
We  presented an implement of the game
of coauthoring behavior
through a geometric hypergraph,     expressing
 the    reputations and   costs     through a function of   hyperdegree  and   that of   geometric distance  respectively.

%Positive    benefit-minus-cost  is the condition of cooperation.

The model  is built   on  a circle $S^1$ as follows.
 Sprinkle   a set  of nodes  $N_t$
on $S^1$   uniformly and randomly at each time $t\in [1,T]$, where $T\in \mathbb{Z}^+$.
  Select a subset  $N^l_t$ from $N_t$ randomly  as lead nodes.
  Each
 lead  node $i$   is assigned an arc with centre $\theta_i$ to    imitate  its    reputation. The length of the  arc is \begin{equation}
r_i(t) = \frac{\alpha}{t}(h_i(t)+1)^{\mathrm{e}^{-\beta(k_i(t)+1)}},
\label{eq1}
\end{equation}
where $h_i(t)$ is the hyperdegree of node $i$ at time $t$,  $\alpha$ and $\beta\in \mathbb{R}^+$.
 The group of the nodes covered by a lead node's   arc  expresses
a   research team.

Coauthoring behavior is expressed through
 hyperedges generated as follows.   For each new node $j(\theta_j,t_j)\in N_t$, select a lead node set $M^l_j$ for which   $\forall i (\theta_i,t_i)\in M^l_j$ satisfies $r(i)>\pi-|\pi-|\theta_{i}-\theta_{j}||$ and $t_i<t_j$. It means node $j$'s reputation is larger than the angular distance between $i$ and $j$,
  an expression of  positive    benefit-minus-cost.
For  each  $i\in M^l_j$, we append $j$ to $R_i$ (a set used to model  node $i$'s  research team),
and  generate a hyperedge at probability $p$ by grouping   $i$, $j$, $(\min(x,|R_i|)-2)$  nodes of $R_i$   nearest to $j$, and
$(x-\min(x,|R_i|))$     nodes  $\not\in R_i$ randomly,  where   $x$ is the random variable  of an inputting distribution,    a fitting of   the empirical distributions of  hyperedge sizes.

The  way of generating hyperedges  describes a usual  scene of cooperation.
A researcher   wants to complete a  work, which needs several researchers to work together.
Then he   would ask  his team leader    for help.  The leader would  suggest some team members
  with similar interest to work together.
Such   behavior can be viewed as kin selection.
When finishing the  work is beyond the ability of  the   team, the researcher   would ask for external helpers. This  inspires the design of randomly choosing  $(x-\min(x, |R_j|)  $  nodes outside of $   R_j $
 to cooperate.

The random variable $x$, used to model hyperedge size, is generated as follows.
Give   the upper bound of small research team  $\mu>0$, and the lower bound   of large research team  $\nu>0$.
 Denote the expected value of hyperedge sizes and the size of
$R_i$   to be $\eta$ and $\lambda$.
Let  $\eta    = \min \{ \lambda  ,\mu\}$ if   $ \lambda \leq\nu $, and
draw $\eta $  from a power law distribution with  an exponent $\gamma$ and   domain $[\mu,\lambda]$ if   $\lambda>\nu $.
Draw
  $x$   from a Poisson distribution with  the expected value $\eta$.

%Note that in the description of the above  game,   $x=s+2$   and $\lambda=m+2$.

%
% emerge  a hook head and a fat tail,
% which means the sizes of substantial papers  are around their average, and a few papers have a  significantly  large size.
% In reality,   researchers in a small research team
% are more likely to write papers
%together. Members of a large research team rarely coauthor a paper all together, but rather  with  a   fraction of members.

We   assumed each lead node has the same attraction to new nodes,  and so let the reputation of  a lead node $i (\theta_{i},t_{i})$  be
     $r( i)\propto 1/t_i $ (which is   inherited from the PSO  model\cite{Krioukov2013}).
       The novel aspect we introduced is
 the fading   heterogeneity  over  hyperdegrees.
 When a node's   zonal size
  fades to a value free of its hyperdegree, the  probability  of receiving connections is the same as that of the standard random   graph,   which   has an
exponentially decaying hyperdegree distribution. In contrast,
if its hyperdegree is not   large enough, then cumulative advantage is significant, which gives rise to the power law
  part of  hyperdegree distributions.

%The size  of  influential zone   is inversely proportional  to time, which
% gives an  expression  of the inefficient  information    of new players.
%

 Compared with the model in Ref.\cite{Xie6}, our model removes the  hypothesis of  the arc-length being a power function of nodes' birth time,
  thus gives a   direct simulation  of the  cumulative advantage for   hyperdegrees and    degrees. It  expresses
  a typically  cooperative mechanism, namely  network reciprocity. In addition,  compared with that previous  model, our  model has the   ability to predict the the transition from  power law to exponential cutoff for     hyperdegree and degree
distributions, thus   fully reproduces the trichotomy of these  distributions.

%Based on
%the cost and the benefit of collaborations, we explain    the  distribution feature   of paper team sizes.
% The   law of diminishing marginal utility holds in academic society.
%The allocation of academic achievements is often according to author order.
%   Hence only the researchers with positive
%  benefit-minus-cost would  collaborate  papers.

\section*{The data}

We analyzed two   empirical coauthorship datasets  collected  from Web of Science (www.webofscience.com).
  Dataset  PNAS  is composed of  36,732     papers published in   {\it Proceedings of the National Academy of
Sciences} during 2007--2015.
Dataset PRE comprises 24,079 papers published in {\it Physical Review E}    during  2007-2016.
The different collaboration level (reflected by the average number of the authors per paper) of the two datasets (PNAS 6.712, PRE 3.102)
helps to test  model flexibility.

 The entities of persons are identified by    authors' name on their  papers, which constitute
 the nodes of the empirical networks analyzed  here.  Table~\ref{tab1} shows certain statistical indexes for these networks.
Note that these networks
suffer errors: one person is identified as two
or more entities (splitting error); two or more persons are identified as one entity (merging error).
The analysis in Appendix A shows    the multimodality phenomena considered here  are robust to these errors at certain levels.

 \begin{table*}[!ht] \centering \caption{{\bf Specific   statistical   indexes of  empirical  networks.} }
\begin{tabular}{l r r r r r r r r r} \hline
Network&NN&NE   & GCC & AC &AP  & PG     \\ \hline
  PNAS  &161,780&1,074,836 &0.896   &0.554 &6.599   & 0.848&    \\
PRE & 37,528 &90,711&  0.838&  0.394 &6.060  &   0.583 \\\hline
 PNAS-s &115,462&1,049,253 & 0.832 &0.064 &4.176  &  0.968\\
 PRE-s   &27,925& 86,648&0.773 & 0.237 &6.148  &0.814 & \\\hline
Synthetic-1 & 103,816 & 613,466  &0.574  & 0.255  & 12.23   &  0.574  \\
Synthetic-2   & 34511 & 73665&  0.826&  0.338 &   13.63  & 0.440 \\
\hline
 \end{tabular}
  \begin{flushleft} The indexes are    the numbers of nodes (NN) and edges (NE),  global  clustering coefficient (GCC),  assortativity coefficient   (AC),  the average shortest path length (AP),    and the node proportion of the giant component~(PG).
  The values of AP  of   the first, third and fifth networks are calculated by sampling 300,000 pairs of nodes. The networks with suffix -s are used in Appendix A.
\end{flushleft}
\label{tab1}
\end{table*}

\begin{figure}\centering
% Use the relevant command to insert your figure file.
% For example, with the graphicx package use
\includegraphics[height=1.5  in,width=6.   in,angle=0]{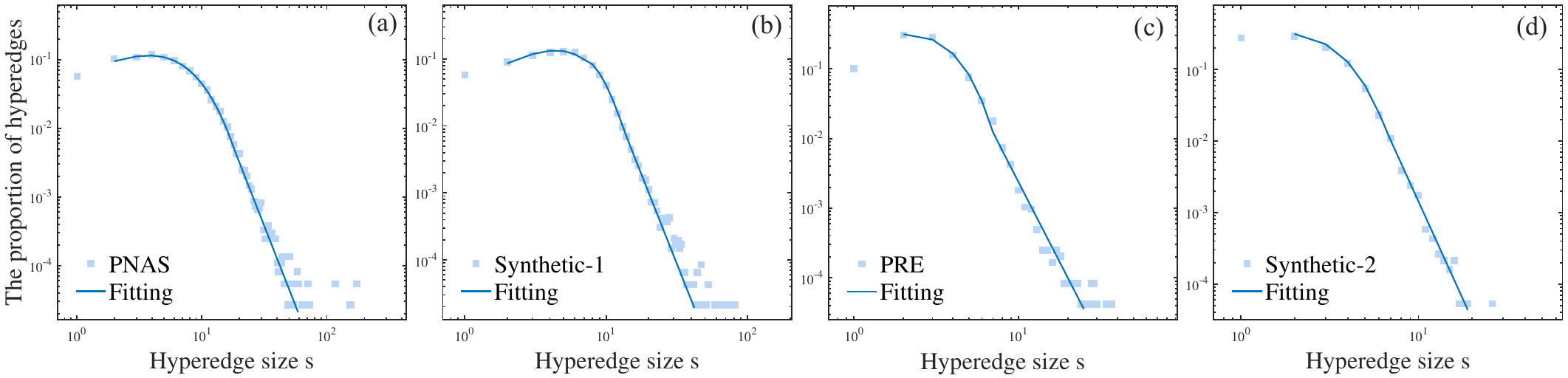}
%\includegraphics[height=1.6  in,width=6.   in,angle=0]{Figure_8.pdf}
% figure caption is below the figure
\caption{{\bf The   distributions   of hyperedge sizes.   }
Each fitting is a mixture  of a generalized Poisson      and a power-law distribution.
 The fitting process and  goodness-of-fit are shown in Appendix B. } \label{fig3}      % Give a unique label Index $p$ is the proportion of the generalized Poisson distribution.
\end{figure}

We modelled two hypergraphs to reproduce   the multimodality phenomena  emerged in the empirical  networks.
The reason   of choosing their parameters (Table~\ref{tab2}) is as follow.
 We valued $\mu$ through the expected value of the general Poisson part of an empirical distribution of hyperedge sizes~(Fig.~\ref{fig3}), and
  valued $\nu$ through  iteration from  the starting point   of the power-law part of the distribution until the distribution of modelled hyperedge sizes is similar to  the empirical
one.
We
 valued    $\alpha$,  $|N_t|$, $|N^l_t|$ and $p$ to make  the hyperdegrees of substantial nodes be one.

 %~(Synthetic-1:  48.5\%, Synthetic-2:  61.5\%).

  \begin{table*}[!ht] \centering \caption{{\bf The parameters of   synthetic networks.} }
\begin{tabular}{l llllllll  } \hline
  $T=6,000, 9,000$ & $N_t=100, 15$&$N^l_t=5,   5$& $p=0.25, 0.4$ & \\
   $\alpha=0.073, 0.19$&   $\gamma=0.001, 0.00001 $& $\mu=6, 2$& $\nu=42, 6$& \\
\hline
 \end{tabular}
  \begin{flushleft}
The parameters in the first row control   network size, and those in the second row control the distribution of hyperedge sizes,  degree
and hyperdegree distribution.
\end{flushleft}
\label{tab2}
\end{table*}

 \section*{The trichotomous   distributions of degrees and   hyperdegrees}
 \label{sec3}

%of the
%cumulative  degree  distributions  of    these coauthorship networks, as well as their hyperdegree distribution.
% Behaviors of collaborating and writing are dependent on the choices of authors, the
%attractiveness of authors and papers. The choices can be simplified by ``yes/no" decisions.
%Take writing  behavior as an example.
%Treat the event that whether writing a paper   as a ``yes/no" decision.

\begin{figure}\centering
% Use the relevant command to insert your figure file.
% For example, with the graphicx package use
\includegraphics[height=3.   in,width=6.    in,angle=0]{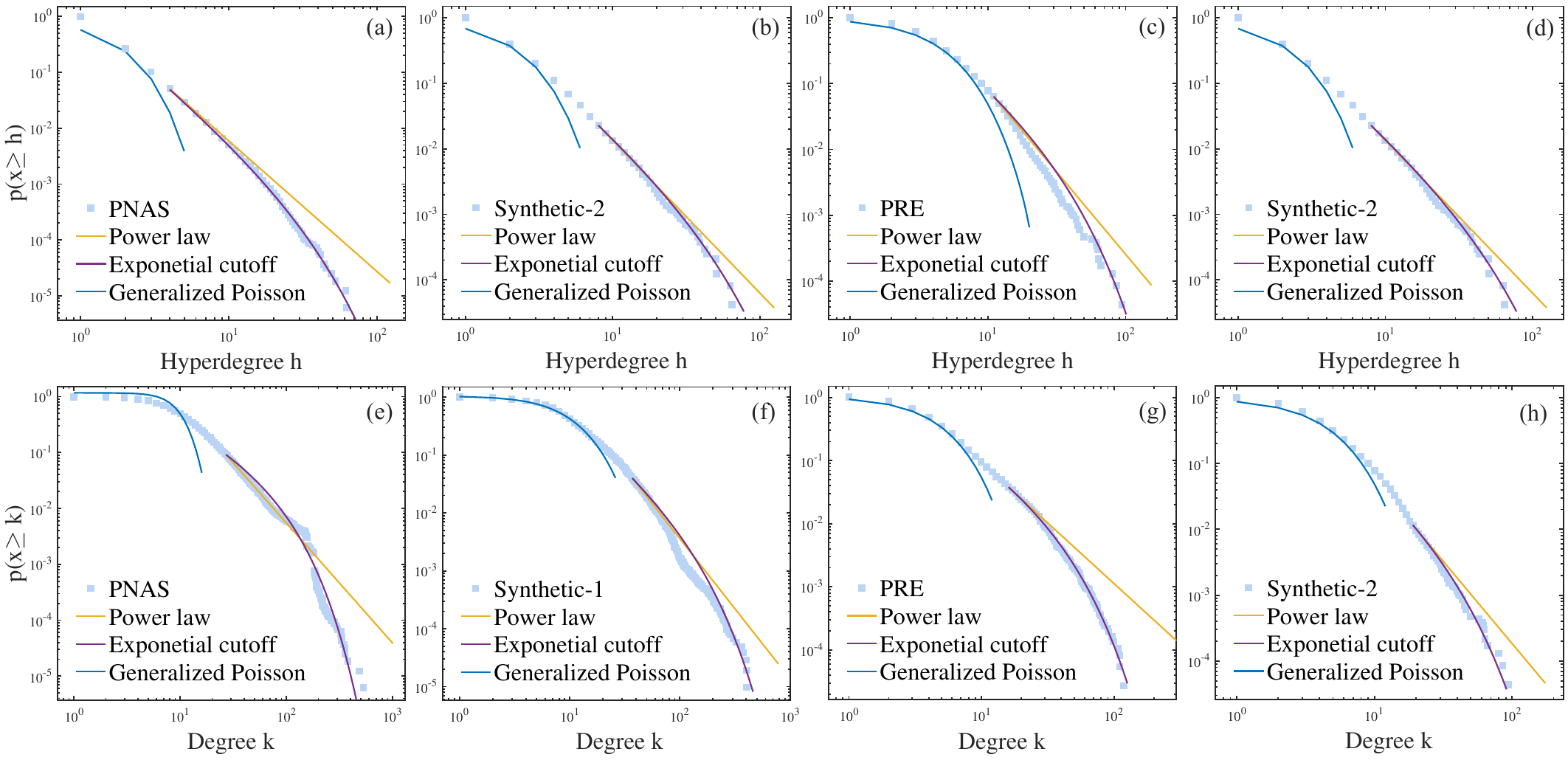}
% figure caption is below the figure
  \caption{ {\bf The cumulative distributions   of   hyperdegrees and degrees.} Panels show these cumulative
distributions  of  the empirical and synthetic datasets, as well as      their  trichotomy.  }
 \label{fig1}      % Give a unique label
\end{figure}

The trichotomy of  degree and  hyperdegree   distributions  comprises a generalized Poisson head, a power-law in the middle part, and an
exponential tail~(Fig.~\ref{fig1}).
 The power law parts are fitted through the  method of Clauset et al\cite{Clauset2009}.
  The generalized Poisson distributions and the   power law  functions   with an exponential cutoff are  fitted through  maximum-likelihood  estimation. The  fitting process and     goodness-of-fit
  are shown    in Appendix C.

Generalized Poisson distributions can be derived from  a range of ``coin flipping" behaviors, where  the    probability of observing ``head"     is dependent on observed events\cite{Consul}.
The event of publishing a paper  is kind of observing ``head" (reject or accept),
where the probability of publishing
  is also affected by previous
events. For example, an  author would easily write his second paper, compared with his first one.
There is a cumulative advantage in writing papers, writing
experiments   accumulating  in the process of publishing papers.
In hyperdegree distributions, the  process
reflects as the transition from  generated Poisson  distribution  to     power law.
Meanwhile, aging  is against cumulative advantage, which reflects as the transition from  power law    to    exponential cutoff.

Fig.~\ref{fig1} shows our model can fully reproduce the trichotomy.
The modelled hyperedges are generated through an inhomogeneous  Poisson  point process (namely its density of points is not a constant), because
the probability  of new nodes fallen   in a lead node's zone   depends on the lead node's hyperdegree.
     At each time $t$, the expected number of new hyperedges    $r_i(t)m/2\pi$  is a function of $h_i(t)$, the  real-time hyperdegree at time $t$.
     Therefore,
the hyperdegree $h_i(T)$    follows  a generalized Poisson distribution.
When $h_i(T)$ is sufficiently large and $\beta(h_i(t)+1)\ll1$, the formula~(\ref{eq1}) gives rise to
\begin{equation}\frac{d}{dt}h_i(t)=\frac{m}{2\pi}r_i(t) \approx     \frac{\lambda}{  t}  (h_i(t)+1)   , \label{eq2}
\end{equation}
  where  $\lambda= {\alpha m}/{2\pi  }$.
The solution to Eq.~(\ref{eq2})
gives node $i$'s expected hyperdegree   $\bar{h}_i(T)=(T/t_i)^{\lambda}-1$, which   gives rise to    $p(\bar{h}_i(T)\leq h) =  p (t_i \geq    T/(h+1)^{1/\lambda}  ) $.
Based on the mean-field theory, we   deduced its density function as follows.
 The probability
 of a  node  generated at $t_i$ is $1/T$. Hence
 $ p (t_i   \geq    T (h+1)^{-1/\lambda} ) =1-p (t_i  <    T (h+1)^ {-1 / \lambda} )=1-   (h+1)^{- {1}/{\lambda}} $, which gives rise to
a power-law distribution
 \begin{equation} p(h)=\frac{d }{d h}p(\tilde{h}_i(T) \leq   h) \propto  (h+1)^{-1-\frac{1}{\lambda}}\approx h^{-1-\frac{1}{\lambda}}\label{eq4}
\end{equation} for  sufficiently large $h$.
It shows
the transition from   generalized Poisson distribution   to    power law.

When $\beta(h_i(t)+1)\gg1$,  the formula~(\ref{eq1}) gives rise to
\begin{equation}\frac{d}{dt}h_i(t)=\frac{m}{2\pi}r_i(t) \approx  \frac{\lambda}{  t}  . \label{eq3}
\end{equation}
The solution to Eq.~(\ref{eq3})
gives node $i$'s expected hyperdegree  $\bar{h}_i(T)= \lambda \log (T/t_i)+\delta  $,
where $\delta$ is the    hyperdegree accumulated from  the process governed by Eq.~(\ref{eq2}).
It   yields   $p(\bar{h}_i(T)\leq h) =
  p (t_i \geq    T \mathrm{e}^{-  (h-\delta)/\lambda}  ) $.
 Hence
 $p (t_i  \geq T \mathrm{e}^{- (h-\delta)/\lambda}  ) =1-p (t_i  <    T \mathrm{e}^{-  (h-\delta)/\lambda} )=1-  \mathrm{e}^{-  (h-\delta)/\lambda}$. It gives rise to \begin{equation}p(h)=\frac{d}{dh} p(\bar{h}_i(T) \leq   h) \propto \frac{1}{\lambda}  \mathrm{e}^{- \frac{h-\delta}{\lambda}} ,\label{eq5}
\end{equation} which  is an exponential distribution on the interval $[\delta,+\infty)$.
 Therefore, we can expect  a   power law with an exponential cutoff  for  sufficiently large  hyperdegrees, namely $
p(h ) \propto  h^{-1- {1}/{\lambda }}\mathrm{e}^{-{h}/{\lambda}}$.
%With the growing of their papers, a few authors   experience  the cumulative process of collaborators over time, whose
%   reputations also increase.
%The process  is often explained by    cumulative advantage.

\begin{figure}\centering
% Use the relevant command to insert your figure file.
% For example, with the graphicx package use
\includegraphics[height=1.5  in,width=6.    in,angle=0]{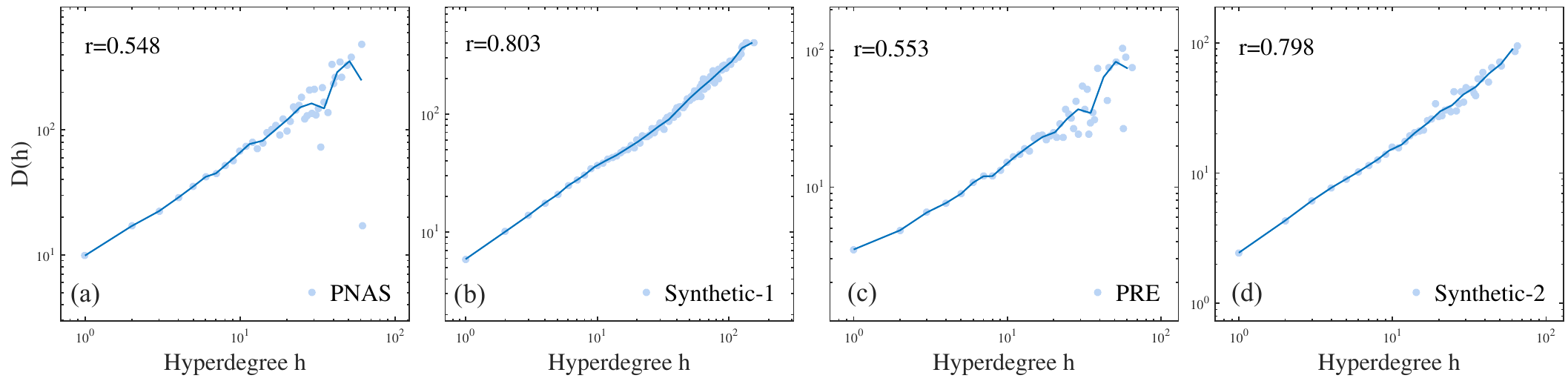}
% figure caption is below the figure
 \caption{{ \bf  The positive correlation  between degrees and hyperdegrees. }
 Function $D(h)$ is  the average  degree of the nodes with hyperdegree $h$.
 Index $r$ is the Pearson's linear correlation coefficient.  Data are binned
on abscissa axes to   improve visibility for the positive slopes.
}
 \label{fig4}      % Give a unique label
\end{figure}

Now we turn to the trichotomy of degree distributions~(Fig.~\ref{fig1}).
The hyperdegrees of many nodes are equal to   one    (PNAS: 74.0\%, PRE: 63.9\%).
As Fig.~\ref{fig3} shows, most of hyperedge sizes follow a generated Poisson distribution (PNAS: 99.9\%,  PRE: 99.9\%).
Those lead the  generalized Poisson parts of degree distributions.
Meanwhile,  nodes' hyperdegree  positively correlates to their degree~(Fig.~\ref{fig4}),
  the growing   process of hyperdegrees coupling  with that of degrees.
The positive correlation leads   the multimodality of degree distributions.
Note that  the nodes of  hyperedges with a large size   also have a large degree,
which leads    the outliers in the tails of degree distributions.

In synthetic data,
The hyperdegrees of a large fraction of   nodes are equal to one~(Synthetic-1: 47.0\%, Synthetic-2: 60.7\%). The   sizes of most   modelled hyperedges follow  a generalized   Poisson distribution~(Synthetic-1: 99.8\%, Synthetic-2: 99.7\%). Those yield the generalized   Poisson part of   modelled degree distributions.
Meanwhile, the
 model preserves  the positive correlation between   hyperdegrees and degrees, especially $h(t)\propto k(t)$ for the nodes with a large hyperdegree.
Therefore, it can reproduce   the multimodality of degree distributions.

\section*{The dichotomous phenomenon  in  clustering and   assortativity}

Coauthorship networks are found to have
two features: high   clustering  (a high probability of a node's two neighbors connecting) and   degree assortativity (nodes' degree positively correlates to their average degree of  neighbors), which are reflected through  the high values of their global clustering coefficient  and
 assortive coefficient~(Table~\ref{tab1}).

\begin{figure}\centering
% Use the relevant command to insert your figure file.
% For example, with the graphicx package use
\includegraphics[height=3.  in,width=6.   in,angle=0]{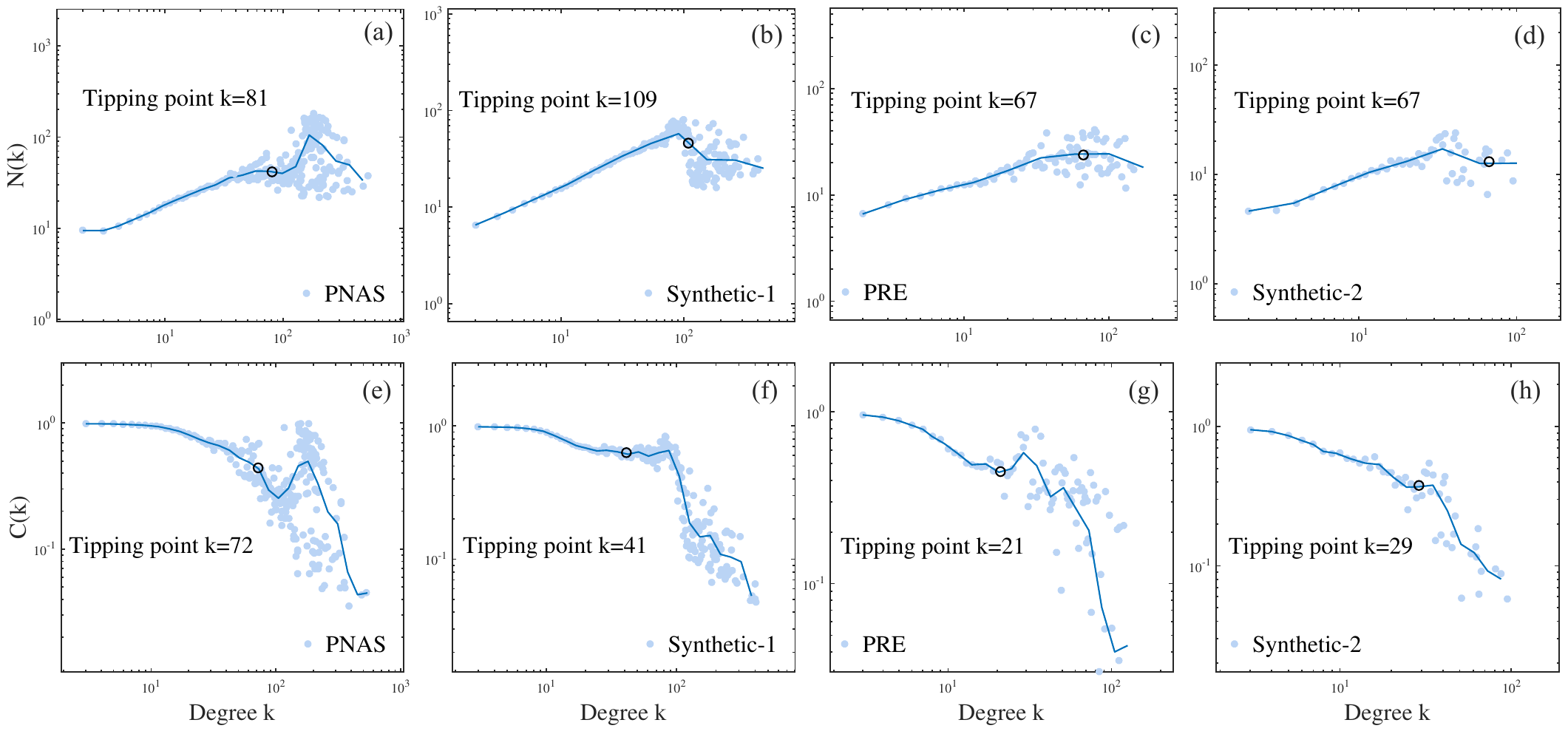}
% figure caption is below the figure
\caption{  { \bf The  average local  clustering coefficient of $k$-degree nodes $C(k)$ and  the average degree of $k$-degree nodes'  neighbors  $N(k)$.}  The tipping points of  the two functions
are detected by the boundary point  detection algorithm for general functions in Ref.\cite{Xie6}, which is listed in Appendix C.
 } \label{fig5}      % Give a unique label
\end{figure}

Observing
these features over degrees, we can find that they  differ from   small degree nodes   to
large degree nodes.
Consider   the  average local  clustering coefficient of $k$-degree nodes $C(k)$ and  the average degree of $k$-degree nodes'  neighbors  $N(k)$.
We can find the  dichotomy of  the functions $C(k)$ and  $N(k)$~(Fig.~\ref{fig5}).

The dichotomy is due to
the positive correlation between hyperdegrees and degrees.
  A large fraction of  nodes with a small degree only belong to  a few hyperedges.
  Their neighbors    probably connect to  each other, and only a few
   of their neighbors have a large degree.
  Therefore, over small $k$,
  the value of $C(k)$ is high,     and the slope of $N(k)$ is positive.
    Consider the  small fraction of nodes  with a large degree.
These nodes   probably have a large hyperdegree.
Their neighbors in different hyperedges  probably do not connect with each other and have a small degree. Otherwise, there will be many nodes with a large degree, out of touch with reality.
  Therefore,  the slopes   of $C(k)$      and   $N(k)$ over large $k$  are negative.
 Our
model   captures   the positive correlation, thus gives a reasonable fit to the dichotomy.

\section*{Discussion and conclusions}

 A   hypergraph model is provided to
simulate the evolution of coauthorship networks from the perspective of    cooperative game, which
expresses two   cooperative mechanisms,   kin selection
and network reciprocity, in a geometric way.
It predicts  a
 range of  features of coauthorship networks, such as fat-tail, small-world, etc.
Especially,
it reproduces the trichotomous distributions for degrees and  hyperdegrees,
as well as the dichotomous phenomena in clustering and  assortativity.
 It overcomes the  weakness of the   model in Ref.\cite{Xie3}, providing a direct simulation of  the cumulative
 advantage on writing experiments.

The model provides an example of how
individual  strategies based on
positive benefit-minus-cost and on specific randomness
generate the   complexity emerged
in coauthorship networks. It  can be extended to analyze the coevolution with citation behavior\cite{Xie7}.
 It also has the potential to be a null model in the empirical analysis
of social affiliation networks.
 Whereas   other typically  cooperative mechanisms should be addressed, such as group selection, direct and indirect reciprocity.

%distributions for citations, interdisciplinary collaborations,

\section*{Acknowledgments}  This work is supported by    National   Science Foundation of China (Grant No. 61773020).

\section*{Appendix A}

 Ambiguities exist in  coauthorship data, known as merging   and splitting errors.
 Since there are many nodes with a small degree or hyperdegree, these  errors  do not change  the feature   of the heads of degree  and hyperdegree distributions, as well as the heads of
     $C(k)$ and $N(k)$.
 Merging errors can ruin  the exponential cutoff, because they generate  nodes with a  degree
far more   than   ground truth.
Splitting errors generate nodes with a degree smaller than ground truth, thus would generate exponential cutoff.

\begin{figure*}[h]
\centering
% Use the relevant command to insert your figure file.
% For example, with the graphicx package use
\includegraphics[height=3.   in,width=6.   in,angle=0]{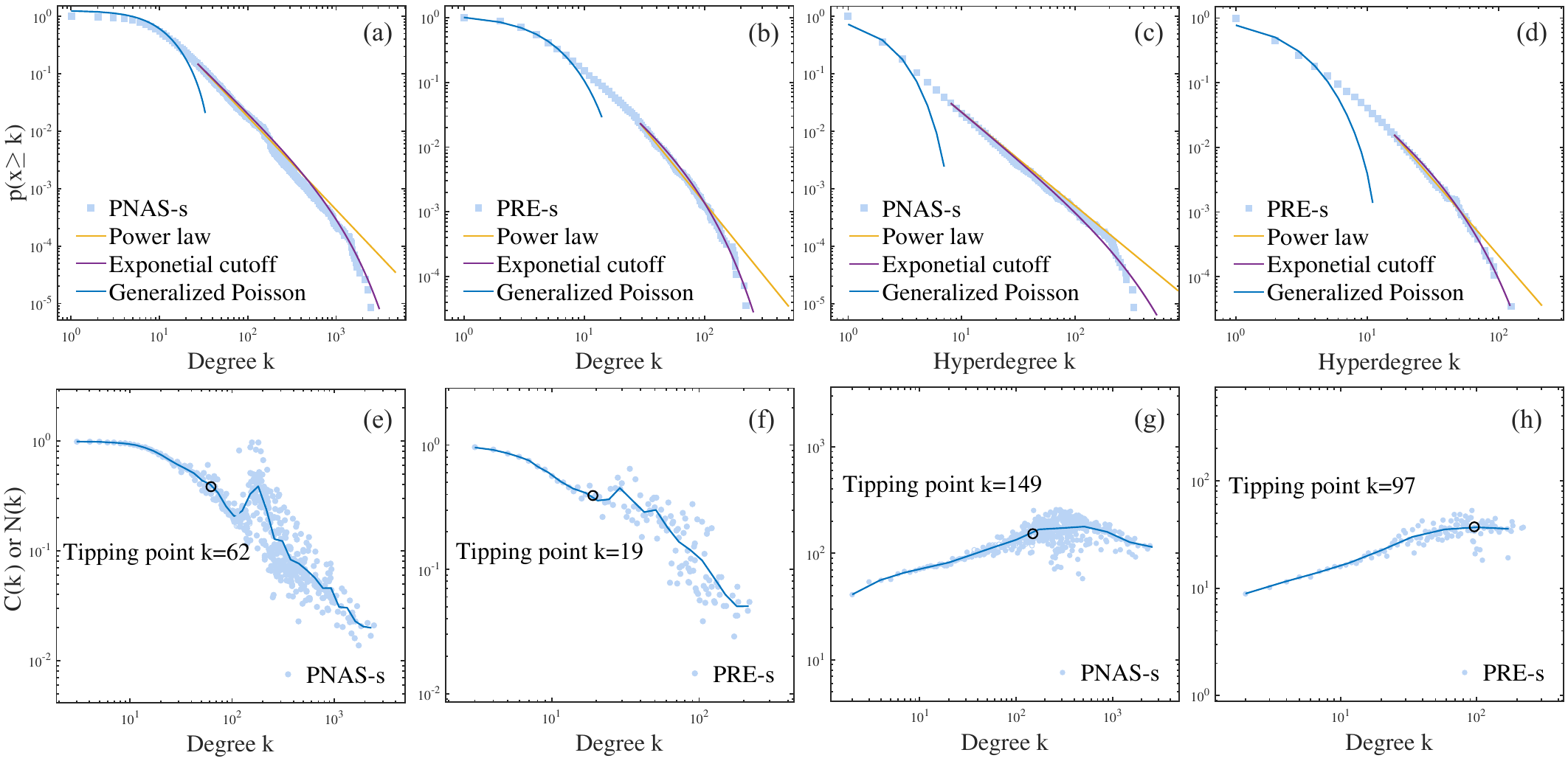}
% figure caption is below the figure
\caption{    {\bf The multimodality phenomena of the networks generated by using short names.}
 Panels show these networks'   degree and hyperdegree distribution, $C(k)$ and $N(k)$.   }
 \label{fig6}
\end{figure*}

Consider   a network suffering heavy merging errors and free of splitting errors.
If the network
  has an exponential cutoff in its degree and hyperdegree distribution,
we could say
these   distributions of the corresponding ground-truth network has
an  exponential cutoff.
In fact,
the density function of
the summation of two random variables drawn from an exponential distribution still has an exponential factor.
 Generate     entities  through
 short name (surname and the initial of the first given name) and use them to construct networks   (PNAS-s, PRE-s).
 These networks are almost free of splitting errors (unless an author provides a wrong name), but heavily suffer merging errors.
 Table~\ref{tab1} shows certain statistical indexes of these networks.
   Fig.~\ref{fig6} shows the degree   and hyperdegree   distribution
  still have an exponential cutoff, which means    the  cutoff exists in the corresponding ground-truth
    distributions.

Merging errors satisfying specific  distribution can   generate  a power law. Therefore,
our conclusions are drawn under the assumption that the ground-truth coauthorship networks have
a power law part in their degree and hyperdegree distribution,
 because the considered data  are not free merging errors. The explanation for  the features of the tails of
     $C(k)$ and $N(k)$ is also dependent on this assumption.

To show the reasonability of this assumption,
 we computed the proportion
of     short names in the names on papers,
and that of the short   names   appearing in  more than one paper, because
  using short names will   generate  a lot of merging errors\cite{Milojevic3}.
 Chinese names were also found to
account for merging errors\cite{Kim1}. For the names on papers, we counted the
proportion of the names  consisting of  a surname
among major 100 Chinese surnames and a given name less than six characters.   The small proportions of such   names
and
those of such names   appearing in  more than one paper
  imply
  the impacts of merging errors are limited, especially for the dataset PNAS~(Table~\ref{tab4}).
\begin{table*}[!ht] \centering \caption{{\bf Specific   statistical   indexes of  the empirical data.} }
%\scriptsize
\begin{tabular}{l r r r r r r r r r r} \hline
Data&     $a$ & $b$ & $c$ & $d$\\ \hline
  PNAS & 1.40\% &0.13\% &  3.11\%&  1.30\%  \\
PRE  &19.2\%&6.45\% &3.85\% &1.58\% \\
%Nature     &  	91.5\% 	&98.6\% &	3.7\% & 1.2\% &86.7\%  &85.6\%\\
%Science & 	 88.9\%	&97.6\%	&3.6\%& 1.0\% &86.8\%  &80.2\%\\
\hline
 \end{tabular}
  \begin{flushleft}
  Indexes $a$ and $b$  are the proportion of   short names in the names on papers,
 and the proportion of  the short names   appearing in more than one paper.
   Indexes $c$ and $d$ are
  the proportion of the names     consisting of  a surname among major 100    Chinese surnames and only one given name shorter than six characters,
 and the proportion of such names     appearing in   more than one paper.

\end{flushleft}
\label{tab4}
\end{table*}

\section*{Appendix B}

The empirical distribution of hyperedge sizes can be fitted by a mixture  of a generalized Poisson distribution   and a power law.
Let
the domains of  generalized Poisson  $f_1(x)= {a (a+bx)^{x-1}}{ \mathrm{e}^{-a-bx }/{ x!}} $,  cross-over and power-law   $f_2(x)=cx^{-d}$  be $[\min(x),E]$,  $[B, E]$ and  $[B,\max(x)]$ respectively.
  The fitting function defined on $[\min(x),\max(x)]$ is $f(x)=q(x)      s f_1(x) +(1-q(x))f_2(x)$, where
  $ q(x)=\mathrm{e}^{ - (x -B)/(E-x ) }$.    The fitting processes are:
  calculate parameters of $sf_1(x)$   and $f_2(x)$  by regressing  the head and tail of  empirical  distribution  respectively;      find  $B$ and $E$ through exhaustion    to  make $f(x)$ pass  the Kolmogorov-Smirnov~(KS)   test ($p$-value$>0.05$).  The fitting results are listed in Table~\ref{tab4}.

\begin{table*}[!ht] \centering \caption{{\bf  The   fitting  of  the distribution  of hyperedge sizes.} }
\begin{tabular}{l rr r r rr r r r r r r r} \hline
  Networks  &$a$ &$ b$   & $c$ &$d$  &$s$ & $B$   & $E$      &$p$-value  \\ \hline
PNAS &     3.504 &  0.444   & 2,086  &4.428&   1.159& 12 &32 &  0.913   \\

Synthetic-1 &  4.406     &  0.234   &  1,919&4.794  &1.056  & 7& 18 & 0.074  \\

PRE &    2.507   & 0.000   &92.00& 4.587 & 1.249 &  6 &7 &0.133    \\

 Synthetic-2 & 2.067 &0.049  & 409.9   & 5.454 &  1.156  &5 &6& 0.651\\  %1473  1.259
\hline
 \end{tabular}
  \begin{flushleft}
The  $p$-value of KS test is a measurement of goodness-of-fit.
   \end{flushleft}
\label{tab4}
\end{table*}

\section*{Appendix C}

Table~\ref{tab5} shows the
  boundary    detection  algorithm for   generalized Poisson distributions.
  The  observations  $\{D_s, s=1,...,n\}$ are nodes' degree  or hyperdegree.
 Changing the algorithm as follows, we can obtain the
  boundary of  the power law functions
with an exponential cutoff. For   $k$ from $1$ to $\max(D_1,...,D_n)$,
we fit   $h(\cdot)$  to   the density $h_0(\cdot)$ of    $\{D_s, s=1,...,n|D_s \geq k\}$,
until    the KS test accepts the null hypothesis.
  Table~\ref{tab6} shows the corresponding goodness-of-fit.

\begin{table*}[!ht] \centering \caption{{\bf A boundary detection algorithm of generalized Poisson distributions\cite{Xie6}.} }
\begin{tabular}{l r r r r r r r r r} \hline
Input: Observations  $D_s$, $s=1,...,n$,  rescaling function $g(\cdot)$, and fitting model     $h(\cdot)$.\\
\hline
For   $k$ from $1$ to $\max(D_1,...,D_n)$ do: \\
~~~~Fit   $h(\cdot)$  to   the PDF $h_0(\cdot)$ of    $\{D_s, s=1,...,n|D_s \leq k\}$    by  maximum-likelihood\\ estimation; \\
~~~~Do    KS test for two    data
     $g(h(t))$ and $g( h_0(t))$, $t=1,...,k$ \\
   with the null hypothesis  they coming from the same continuous distribution;\\
~~~~Break  if  the test rejects the null hypothesis  at     significance level $5\%$. \\ \hline
Output: The current $k$ as the   boundary point. \\ \hline
 \end{tabular}
    \begin{flushleft}
    \end{flushleft}
\label{tab5}
\end{table*}

%For   $k$ from $1$ to $\max(D_1,...,D_n)$,  do     KS test for two    data
%     $g(h(t))$ and $g( h_0(t))$, $t=1,...,k$  by linear regression.
%Break  if  the test accepts the null hypothesis  at     significance level $5\%$.

\begin{table*}[!ht] \centering \caption{{\bf  The   goodness-of-fit of degree and hyperdegree distributions.} }
\begin{tabular}{l  rr r r| rr r r r r r r r} \hline
 Network   &$p_1$ & $x_{1}$  &$ p_2$  & $x_2$  &$p_3$  & $x_3$ &$p_4$ & $x_4$ \\ \hline
PNAS &   0.950 &  3   & 0.538 & 3 &    0.453 &  13& 0.077 &  347   \\
  PRE &   0.240  &  5 &  0.931 &3 &    0.253 &5 & 0.129 & 91   \\
PNAS-s & 0.107   & 5  & 0.975& 3 &    0.656& 14 & 0.112 & 1,476    \\

  PRE-s &  0.166  &14  & 0.320 & 3 &    0.104 &5  &0.077  & 174      \\
  Synthetic-1 &  0.251 &   27 & 0.087 &11 &   0.081& 10 &0.108  &371    \\
  Synthetic-2 &  0.111 & 17 &0.368  &  6&    0.346 & 6 & 0.077 & 81     \\
\hline
 \end{tabular}
  \begin{flushleft}
Indexes $p_1$ and $p_2$ are  the $p$-values of KS test,   measuring the goodness of fitting the heads of degree and hyperdegree distributions by  generalized Poisson distributions.
Indexes $x_1$ and $x_2$ are the right bounds of the corresponding  fitting distributions.
 Indexes $p_3$ and $p_3$ measure    the goodness of fitting   the tails by   power law functions  with an exponential  cutoff.
 The values of  $x_3$ and $x_4$ are the left bounds of the corresponding  fitting   functions.
   \end{flushleft}
\label{tab6}
\end{table*}

Table~\ref{tab7} shows the
  boundary    detection  algorithm for  the  average local  clustering coefficient of $k$-degree nodes $C(k)$ and  the average degree of $k$-degree nodes'  neighbors  $N(k)$.
The inputs are $g(\cdot)=\log(\cdot)$, $h(s)=a_1 \mathrm{e}^{-((s-a_2)/a_3)^2}$ for  $C(k)$, and  $h(s)= a_1 s^3 + a_2 s^2 + a_3 s  + a_4$  for  $N(k)$,  where $s$, $a_i\in  {\mathbb{R}}$ ($i=1,...,4$).
  Using  these inputs is based on   the observations on $C(k)$ and $N(k)$.

\begin{table*}[!ht] \centering \caption{{\bf Boundary point  detection algorithm for $C(k)$ and $N(k)$\cite{Xie6}.} }
\begin{tabular}{l r r r r r r r r r} \hline
Input: Data vector  $h_0(s)$, $s=1,...,K$, rescaling funtion $g(\cdot)$, and fitting model     $h(\cdot)$.\\
\hline
For   $k$ from $1$ to $K$ do: \\
~~~~Fit   $h(\cdot)$  to    $h_0(s)$, $s=1,...,k$  by regression; \\
~~~~Do  KS test for two    data vectors
     $g(h(s))$ and $g( h_0(s))$, $s=1,...,k$ with   the null\\
 hypothesis they coming from the same continuous distribution;\\
~~~~Break  if  the test rejects the null hypothesis  at   significance level  $5\%$. \\ \hline
Output: The current $k$ as the  boundary point. \\ \hline
 \end{tabular}
\label{tab7}
\end{table*}


\begin{thebibliography}{1}


\bibitem{Glanzel1}Gl\"anzel W,  Schubert A (2004)
Analysing scientific networks through co-authorship.
 Handbook of quantitative science and technology research  11: 257-276.

\bibitem{Glanzel2}Gl\"anzel  W (2014) Analysis of co-authorship patterns at the individual level. Transinformacao 26: 229-238.

\bibitem {Sarigol}
Sarig\"ol  E, Pfitzner  R, Scholtes  I, Garas  A,   Schweitzer  F  (2014)  Predicting scientific success based on coauthorship networks. EPJ Data Science  3(1): 1-16.

%\bibitem {Bertsimas}
%Bertsimas  D, Brynjolfsson  E, Reichman  S,   Silberholz  JM  (2014)  Moneyball for academics: Network analysis for predicting research impact. Available at SSRN 2374581.

\bibitem{Mali} Mali F,  Kronegger L,   Doreian P,     Ferligoj A, Dynamic scientific coauthorship networks (2012) In: Scharnhorst A, B\"orner K,
 Besselaar PVD editors. Models of science dynamics. Springer. pp. 195-232.

%\bibitem {Jia}
%Jia T, Wang  D,   Szymanski  BK  (2017)   Quantifying patterns of research-interest evolution. Nature Human Behaviour  1: 0078.

\bibitem {Newman2} Newman M  (2001) Scientific collaboration networks. I. network construction and fundamental
results.  Phys  Rev E 64: 016131.

\bibitem {Newman3} Newman M  (2001)
Scientific collaboration networks. II. shortest paths, weighted networks, and centrality.
 Phys  Rev E 64: 016132.

\bibitem {Newman1}  Newman M  (2001) The structure of scientific collaboration networks. Proc  Natl  Acad  Sci  USA 98: 404-409.


\bibitem {Newman0} Newman M   (2004) Coauthorship networks and patterns of
scientific collaboration. Proc  Natl  Acad  Sci  USA  101: 5200-5205.


\bibitem {Newman4} Newman M  (2002) Assortative mixing in networks. Phys  Rev  Lett  89: 208701.

\bibitem{XieLL}
Xie  Z, Li  M, Li  JP, Duan  XJ,   Ouyang  ZZ  (2018)  Feature analysis of multidisciplinary scientific collaboration patterns based on pnas. EPJ Data Science  7: 5.





\bibitem {Barab} Barab\'asi AL, Jeong H, N\'eda Z, Ravasz E, Schubert A, Vicsek  T. (2002) Evolution of the social network
of scientific collaborations. Physica A  311: 590-614.





\bibitem{Moody} Moody J (2004) The strucutre of a social science collaboration network: disciplinery cohesion form 1963 to 1999. Am Sociol Rev 69(2): 213-238.



\bibitem{Perc} Perc C (2010) Growth and structure of Slovenia's scientific collaboration network. J Informetr 4: 475-482. %(4)

\bibitem{Wagner} Wagner CS, Leydesdorff L (2005) Network structure, self-organization, and the growth of international collaboration in science. Res Policy 34(10): 1608-1618.

\bibitem{Tomassini} Tomassini M, Luthi L (2007) Empirical analysis of the evolution of a scientific collaboration network. Physica A 285: 750-764. %(2)


%\bibitem {Zhou}  Zhou T,  Wang BH,  Jin YD, He DR,  Zhang PP,  He Y, et al. (2007) Modeling collaboration networks based on nonlinear preferential attachment. Int  J  Mod  Phys  C 18: 297-314.

\bibitem{Santos} Santos FC, Pacheco JM (2005) Scale-free networks provide a unifying framework for
the emergence of cooperation. Phys Rev Lett 95(9): 098104.




\bibitem {Catanzaro} Catanzaro M, Caldarelli G, Pietronero L (2004) Assortative model for social networks. Phys  Rev E 70: 037101.


 \bibitem{Nowak} Nowak MA (2006) Five rules for the evolution of cooperation. Science 314(5805): 1560-3.


\bibitem {Krioukov1} Krioukov  D,  Kitsak M,    Sinkovits RS, Rideout D,   Meyer D,   Bogu$\mathrm{\tilde{n}}$\'a  M (2012)  Network cosmology.  Sci  Rep 2: 793.

%\bibitem {Xie1}{   Xie Z,    Rogers T}  (2016)      {Scale-invariant geometric random graphs}. Phys  Rev  E 93: 032310.

%\bibitem {Xie2}Xie Z, Zhu J, Kong DX,  Li JP (2015)  A random geometric graph built on a time-varying Riemannian manifold. Physica A 436: 492-498.

\bibitem {Xie3}{   Xie Z, Ouyang ZZ, Li JP} (2016)    { A   geometric graph   model  for   coauthorship networks}. J Informetr 10: 299-311.	

\bibitem {Xie6}Xie  Z, Ouyang ZZ,   Li  JP,  Dong  EM, Yi  DY (2018)  Modelling transition phenomena of scientific coauthorship networks. J Assoc Inf Sci Technol   69(2): 305-317.




\bibitem   {Krioukov2013}
 {Papadopoulos  F, Kitsak M,    Serrano MA, Bogu$\mathrm{\tilde{n}}$\'a  M, Krioukov  D} (2012)  Popularity versus similarity in growing networks. { Nature}  {489}: {537-540}.


\bibitem{Clauset2009}
Clauset A, Shalizi CR, Newman MEJ (2009) Power-law distributions in emprical data. SIAM Rev 51:
661-703.
\bibitem{Consul}Consul  PC,   Jain GC (1973)  A generalization of the Poisson distribution.
Technometrics 15(4): 791-799.




%\bibitem{Xie9}
%Xie Z, Li JP, Li M, Yi DY, Feng  YQ, Xie ZL (2017)
%Merging error analysis  of name disambiguation based on author similarity

\bibitem {Xie7}{   Xie Z, Xie ZL, Li M, Li JP, Yi DY} (2017)   Modeling the coevolution between citations and coauthorship of scientific papers.
  Scientometrics 112: 483-507.

\bibitem {Milojevic3}Milojevi\'c S (2010) Modes of collaboration in modern science-beyond power laws
and preferential attachment. J Assoc Inf Sci Technol 61(7): 1410-1423.

\bibitem {Kim1}  Kim  J,   Diesner J (2016)  Distortive effects of initial-based name disambiguation on measurements of large-scale coauthorship networks.
J Assoc Inf Sci Technol  67(6): 1446-1461.


%\bibitem{Borner3}
%    B\"orner K, et al. (2010) A multi-level systems perspective for the science of team science. Sci Transl Med 2(49): 49cm24

%\bibitem{Jones}
%Jones BF, Wuchty S, Uzzi B (2008) Multi-university research teams: Shifting impact,
%geography, and stratification in science. Science 322(5905): 1259-1262.
%
%\bibitem{Adams}   Adams J (2012)
%Collaborations: The rise of research networks. Nature  490: 335-336.
%
%\bibitem{Shrum} Shrum W, Genuth J, Chompalov I (2007) Structures of Scientific Collaboration (MIT,
%Cambridge, MA)
%
%\bibitem{Uzzi}  Uzzi B, Mukherjee S, Stringer M, Jones B (2013) Atypical combinations and scientific
%impact. Science 342(6157): 468-472.
%\bibitem{Wuchty}
%    Wuchty S,  Jones BF, Uzzi B (2007) The increasing dominance of teams in production of
%knowledge. Science 316(5827): 1036-1039.




%
%
%
%
%
%
%\bibitem{sci125}Pennisi  E (2005)  How did cooperative behavior evolve?  Science  309(5731): 93-93.
%
%
%\bibitem {Hoekmana} Hoekman J,   Frenken  K,   Tijssen RJW (2010) Research collaboration at a distance: Changing spatial patterns of scientific
%collaboration within Europe. Res Policy  39:  662-673.
%
%
%\bibitem{Hauert}
%Hauert C, Doebeli M (2004) Spatial structure often inhibits the evolution of cooperation in the snowdrift game. Nature  428(6983): 643-6.
%
%
%%\bibitem  {Milojevic5}Milojevi\'c S (2010) Modes of collaboration in modern science-beyond power laws
%%and preferential attachment. J Am Soc Inf Sci Technol 61(7): 1410-1423.
%\bibitem{Milojevic}Milojevi\'c  S  (2014)  Principles of scientific research team formation and evolution. Proc  Natl  Acad  Sci  USA 111: 3984-3989. %(11),
%\bibitem{Borner}B\"orner  K, Maru  JT,   Goldstone  RL  (2004)  The simultaneous evolution of author and paper networks. Proc  Natl  Acad  Sci  USA  101(suppl 1), 5266-5273.
%
%\bibitem {Xie7}{   Xie Z, Xie ZL, Li M, Li JP, Yi DY} (2017)   Modeling the coevolution between citations and coauthorship of scientific papers.
%  Scientometrics 112: 483-507.
%
%
%\bibitem{GoldschmidtC}
%Goldschmidt C  (2005)  Critical random hypergraphs: the emergence of a giant set of identifiable vertices. Ann Probab  33(4), 1573-1600.
%%The Annals of Probability
%
%\bibitem{Estrada2}
% Estrada E, Rodr\'iguez-Vel\'azquez JA (2006)
%Subgraph centrality and clustering in complex hyper-networks.
% Physica A 364,  581-594.
%
%\bibitem{Darling}
% Darling  RW,   Norris  JR  (2005)  Structure of large random hypergraphs. Ann Appl Probab, 15(1A), 125-152.
%
%
%








%\bibitem{Penrose} {  Penrose M}, {\em Random geometric
%graphs}, Oxford studies in probability, 2003.
%\bibitem {Xie}Xie Z, Ouyang ZZ, Zhang PY, Yi DY, Kong DX  (2015)  Modeling the citation network by network cosmology.  Plos One   10(3): e0120687.

%\bibitem {Xie5}Xie Z, Ouyang  ZZ, Liu Q,   Li  JP (2016)  A geometric graph model for citation networks of exponentially growing scientific papers. Physica A  456: 167-175.









%\bibitem{Glanzel}Gl\"anzel  W (2011)
%National characteristics in international scientific co-authorship relations.
%Scientometrics 51:  69-115. % (1)






%
%\bibitem{McPherson}
%McPherson M, Smith-Lovin L, Cook JM (2001) Birds of a feather: homophily in social
%networks. Annu Rev Sociol 27(1): 415-444.
%




%The Annals of Applied Probability


\end{thebibliography}
\end{document}